\documentclass[letterpaper, 10 pt, conference]{ieeeconf}  

\IEEEoverridecommandlockouts                              

\usepackage{amsmath} 
\usepackage{txfonts}
\usepackage{multirow}
\usepackage{booktabs}
\usepackage{graphicx}
\usepackage{pgfplots}
\usepackage{subfig}
\usepackage{lipsum}

\DeclareMathOperator{\indexfunc}{index}
\DeclareMathOperator{\euler}{euler}
\DeclareMathOperator{\vel}{vel}
\DeclareMathOperator{\leftfunc}{l}
\DeclareMathOperator{\rightfunc}{r}

\title{\LARGE \bf
A Unified Editing Method for Co-Speech Gesture Generation via Diffusion Inversion
}

\author{Zeyu Zhao$^{1}$, Nan Gao$^{2}$, Zhi Zeng$^{3}$, Guixuan Zhang$^{4}$, Jie Liu$^{4}$, and Shuwu Zhang$^{3}$
\thanks{*This work was supported by the National Key R\&D Program of China (2022YFF0902202).}
\thanks{$^{1}$Zeyu Zhao is with Institute of Automation, Chinese Academy of Sciences, Beijing, China, and also University of Chinese Academy of Sciences, Beijing, China 
    {\tt\small zhaozeyu2019@ia.ac.cn}}%
\thanks{$^{2}$Nan Gao is with Institute of Automation, Chinese Academy of Sciences, Beijing, China
    {\tt\small nan.gao@ia.ac.cn}}%
\thanks{$^{3}$Zhi Zeng and Shuwu Zhang are with Beijing University of Post and Telecommunications, Beijing, China
    {\tt\small zhi.zeng@bupt.edu.cn}; {\tt\small shuwu.zhang@bupt.edu.cn}}%
\thanks{$^{4}$Guixuan Zhang and Jie Liu are with Beijing University of Post and Telecommunications, Beijing, China, and also Institute of Automation, Chinese Academy of Sciences, Beijing, China
    {\tt\small guixuan.zhang@bupt.edu.cn}; {\tt\small jie.liu@bupt.edu.cn}}%
}

\begin{document}

\maketitle
\thispagestyle{empty}
\pagestyle{empty}

\begin{abstract}

Diffusion models have shown great success in generating high-quality co-speech gestures for interactive humanoid robots or digital avatars from noisy input with the speech audio or text as conditions.
However, they rarely focus on providing rich editing capabilities for content creators other than high-level specialized measures like style conditioning.
To resolve this, we propose a unified framework utilizing diffusion inversion that enables multi-level editing capabilities for co-speech gesture generation without re-training. 
The method takes advantage of two key capabilities of invertible diffusion models. 
The first is that through inversion, we can reconstruct the intermediate noise from gestures and regenerate new gestures from the noise. This can be used to obtain gestures with high-level similarities to the original gestures for different speech conditions. 
The second is that this reconstruction reduces activation caching requirements during gradient calculation, making the direct optimization on input noises possible on current hardware with limited memory. With different loss functions designed for, e.g., joint rotation or velocity, we can control various low-level details by automatically tweaking the input noises through optimization. 
Extensive experiments on multiple use cases show that this framework succeeds in unifying high-level and low-level co-speech gesture editing.

\end{abstract}

\section{INTRODUCTION}

Generating co-speech gestures is a crucial yet challenging task for developing interactive humanoid robots or digital avatars since real human beings often talk with body movements that are hard to precisely predict.
By generating from noisy input with the speech audio or text as conditions, diffusion models are emerging to become the mainstream data-driven methods to produce high-quality synchronizing gestures \cite{zhu2023taming, alexanderson2023listen, ao2023gesturediffuclip, yang2023diffusestylegesture+, ng2024audio}.
Despite the increased quality of the generated gestures, they are still not widely accepted in production environments due to the lack of easy editing measures.
Other than high-level specialized measures like conditioning on gesture styles in different emotion, speaker identity, etc. \cite{10.1145/3414685.3417838,liu2022beat}, a large amount of low-level manual labor including refining the details frame-wise is required for content creators to deliver diverse, meaningful, and stable co-speech gestures.

\begin{figure}[t]
    \centering
    \includegraphics[scale=1.0]{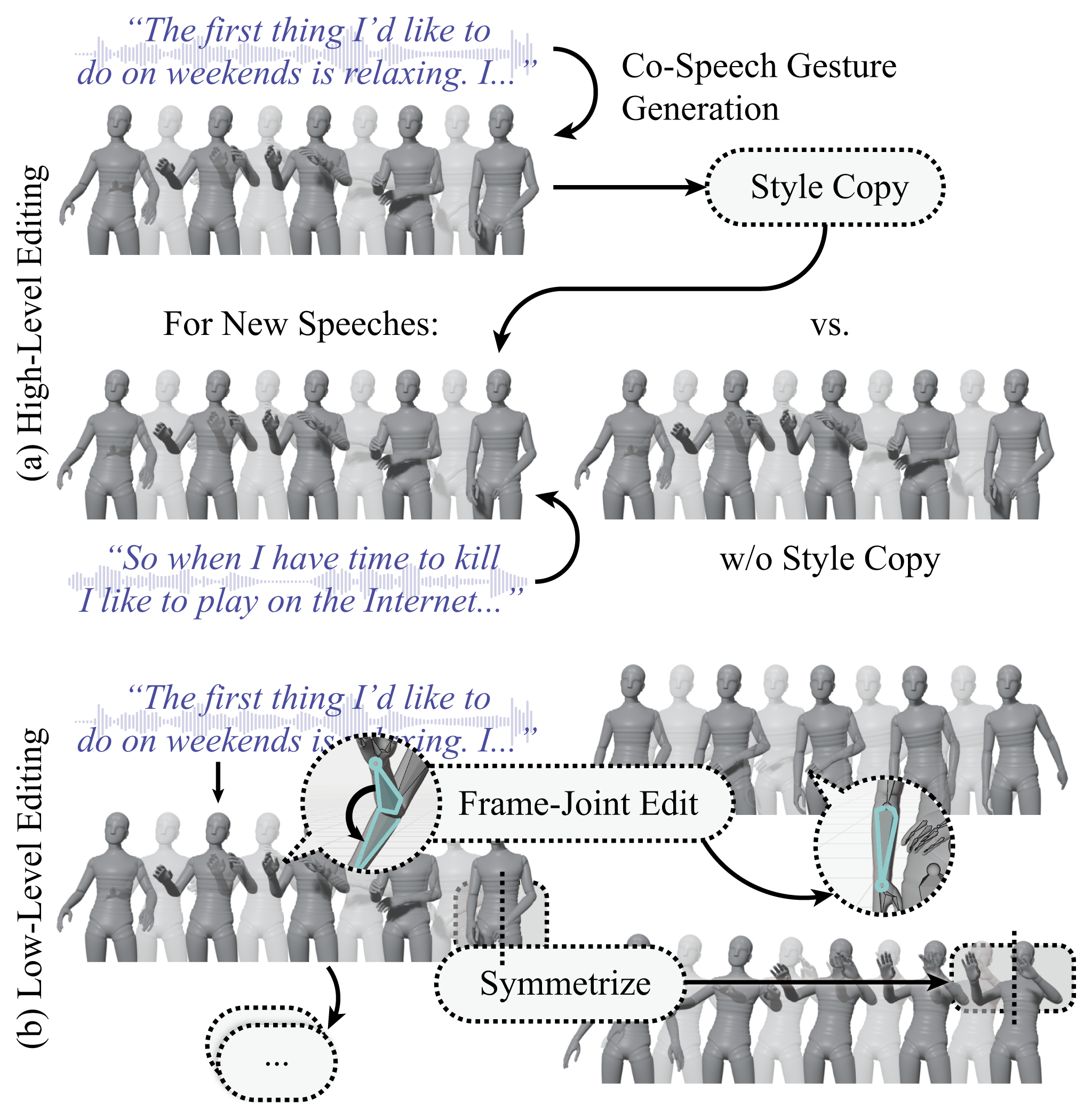}
    \caption{Examples of (a) high-level editing: copying basic style for different speech, and (b) low-level editing: tweaking joint rotation in specified frames or symmetrizing left and right part of the body, etc.}
    \label{fig:examples}
\end{figure}

For a diffusion model, the variation of the input noise is the key factor to sample different target results from the learned distribution lied in the generation process. 
Manipulating this noise can be an easy but powerful method to edit the generated content, in addition to the common but limited measure of conditional generation.
Whether it is to keep the original noise for different conditions, or to automatically tweak the noise in a guided manner, noise editing can provide new possibilities for both high-level and low-level editing, as shown in Fig. \ref{fig:examples}.
However, the original noisy input cannot be directly reconstructed from the generated results due to the one-directional design of the generation process of a regular diffusion model. 
Diffusion inversion \cite{wallace2023edict} is proposed to tackle this problem by enabling the exact inversion of the generation process of regular diffusion models, enabling rich editing capabilities for generated image editing.
Inspired by this, we leverage two key capabilities of the diffusion inversion to provide a unified method of multi-level editing for co-speech gesture generation.
Unlike other means of editing e.g. designing new conditions, it requires no modification to the network structure and does not require any partial or full re-training of the model.
More specifically, our main contributions can be summarized as follows:

\begin{itemize}

    \item By leveraging the intermediate noise reconstruction capability of the diffusion inversion, we demonstrate high-level editing for co-speech gesture generation by reconstructing the original noisy input from the generated gestures and regenerate new gestures from the noise with new conditions to obtain gestures with high-level similarities to the original gestures for different speech.
    \item By leveraging the input noise optimization capability of the diffusion inversion, we also demonstrate low-level editing for co-speech gesture generation by directly optimizing input noises on current hardware with limited memory to control various low-level details by automatically guiding the input noises by a variety of losses.
    \item We design multiple high-level and low-level editing applications and conduct extensive subjective and objective experiments to show that this framework successfully unifies high-level and low-level co-speech gesture generation editing.

\end{itemize}

\section{RELATED WORKS}

\subsection{Co-Speech Gesture Generation}

After early attempts with rule-based methods \cite{10.1145/2485895.2485900} or simple probability modeling methods \cite{10.1145/1778765.1778861}, data-driven methods based on deep learning \cite{nyatsanga2023comprehensive} has become the mainstream for co-speech gesture generation.
Models based on generative adversarial networks (GAN) \cite{ginosar2019learning, 10.1145/3414685.3417838}, variational autoencoders (VAE) \cite{li2021audio2gestures}, normalizing flows \cite{ye2022audio}, and diffusion models \cite{zhu2023taming, yang2023diffusestylegesture+} have shown great success in generating gestures that synchronize well with the input speech audio and text with different style conditioning.
Instead of full end-to-end settings using common generative networks, some models are also utilizing better network structures or modeling designs such as hierarchical decoding \cite{liu2022learning}, phase manifold learning \cite{yang2023qpgesture}, or rhythmic tokenization \cite{ao2022rhythmic} that capture distinct features of the speech input, the gesture output, and their connections.
Some models also use auxiliary modules like contrastive-learning-based text encoders \cite{ao2023gesturediffuclip}, memory networks \cite{9898573} or graphs \cite{zhou2022gesturemaster, zhou2022audio, 10.1145/3577190.3616118} to enhance the generation process to improve certain aspects of the performance in limited conditions.
Most recently, diffusion models have become the first choice for the task \cite{kucherenko2023genea} due to the better modeling capability in theory and the improved quality of generated gestures in practice.
This also allows the expansion of generated modality from co-speech gestures only to more general or holistic manners, like including dancing motions with music \cite{alexanderson2023listen}, or mouth movements and facial expressions \cite{ng2024audio}.

\subsection{Generated Content Editing}

Current artificial-intelligence-generated content (AIGC) still requires extensive manual labor to be post-processed for real production use.
The ability to automatically and easily edit these generated results is essential but under development for content creators who utilize such systems.
Basic editing capabilities enabled by conditional generation \cite{mirza2014conditional} are powerful but require modification and re-training of the model as the conditions are introduced as direct input to the generative model.
In the era of large multimodal models (LMM), partial updating methods like LoRA \cite{hu2021lora} or ControlNet \cite{zhang2023adding} are popular and successful practices to enable more flexible editing capabilities for fine-tuning-required applications.
Another path is to manipulate the latent space of the generative models e.g. by perform guided tweaking to the input noise or the latent output of the input encoders. 
DragGAN \cite{pan2023drag} proposes loss function and tracking techniques that allow interactive editing of GAN-generated images. 
DOODL \cite{wallace2023end} with its image-oriented losses are developed to enable generated image editing for diffusion models in a memory-efficient manner by utilizing diffusion inversion \cite{wallace2023edict}.

Most recently, we noticed that a similar method to ours \cite{karunratanakul2023optimizing} is proposed to enable easy editing capabilities of text to motion generation models. 
This task is very different from ours in that the co-speech gestures are much more nondeterministic and diverse than text-described motions.
This application is also much more time-consuming compared to ours and fails to provide near real-time experiences for end users.
Detailed comparisons can be found in section \ref{sec:editing_latency}.

\section{METHOD} \label{sec:method}

\subsection{Diffusion Models} 

Diffusion models, initially inspired by the diffusion process in non-equilibrium thermodynamics, are generative models that can be trained to sample from a complex probability distribution by reversing the diffusion process that destroys the structure in a data distribution to restore the original structure from noisy input \cite{sohl2015deep}.
Modern Denoising Diffusion Probability Models (DDPM) use neural networks designed for denoising tasks to perform the restoration process and generate novel samples from noisy input in a series of denoising steps \cite{ho2020denoising}. 
Specifically, a neural network $ \epsilon(\mathbf{x}_t, t) $ is designed here to take the noisy input $ \mathbf{x}_t $ at step $ t \in \{T, \dots, 0\} $ and try to predict the noise that is added to the original data in the hypothetical diffusion process. 
Normally, the predicted noise is not directly used to restore the original data. Instead, it is utilized to sample new noisy input for the next step $ t - 1$: 
\begin{equation} 
    \mathbf{x}_{t - 1} = \frac{1}{\sqrt{\alpha_t}} \mathbf{x}_t - \frac{1 - \alpha_t}{\sqrt{\alpha_t (1 - \bar{\alpha}_t)}} \epsilon(\mathbf{x}_t, t) + \sigma_t \mathbf{z},
\end{equation}
where $ \bar{\alpha}_t = \prod_{s = 1}^{t} \alpha_s $, $ \alpha_t = 1 - \beta_t $, $ \beta_1, \dots, \beta_T $ are from a manually defined variance schedule, $ \sigma_t^2 = \frac{1 - \bar{\alpha}_{t - 1}}{1 - \bar{\alpha}_t} \beta_t $, and $ \mathbf{z} \sim \mathcal{N}(\mathbf{0}, \mathbf{I})$.
We start the denoising steps at step $ T $ and finally take $ \mathbf{x}_0 $ as the generated result.
Note that this is not the only sampling strategy for producing $ \mathbf{x}_{t - 1} $. For example, many experiments suggest that directly predicting $ \mathbf{x}_0 $ in each step instead of the added noise can yield better generation in practice \cite{yang2023diffusestylegesture+}. In such cases, the new noisy input for the $ t - 1 $ step can be written as:
\begin{equation} 
    \label{eq:x_t-1_equals}
    \mathbf{x}_{t - 1} = \frac{(1 - \bar{\alpha}_{t - 1}) \sqrt{\alpha_t}}{1 - \bar{\alpha}_t} \mathbf{x}_t + \frac{\beta_t \sqrt{\bar{\alpha}_{t - 1}}}{1 - \bar{\alpha}_t} \epsilon(\mathbf{x}_t, t) + \sigma_t \mathbf{z}.
\end{equation}

\subsection{Diffusion Inversion}

It is clear that by definition, the generation process of DDPMs simulates a Markov chain where each step requires the knowledge of the previous step. 
This is unlike the Denoising Diffusion Implicit Models (DDIM) \cite{song2020denoising}, where the linearization assumption gives $ \epsilon(\mathbf{x}_t, t) \approx \epsilon(\mathbf{x}_{t - 1}, t) $ that makes the reconstruction from $ \mathbf{x}_{t - 1} $ to $ \mathbf{x}_t $ possible.
This inversion capability of DDIMs enables many editing applications by regenerate using the same noise \cite{wallace2023edict} with modified conditions.
However, such inversion based on approximation can be unstable \cite{hertz2022prompt}. More importantly, as a more general and simpler framework, a vast number of applications utilizing diffusion models are still based on DDPMs, especially in the field of co-speech gesture generation \cite{alexanderson2023listen,yang2023diffusestylegesture+}.
Inspired by this, the Exact Diffusion Inversion via Coupled Transformations (EDICT) \cite{wallace2023edict} is proposed to invert the diffusion process of DDPMs and enable similar editing capabilities for general diffusion models, which is nicely suited for our field.

\subsubsection{Inverted Sampling} 

Taking (\ref{eq:x_t-1_equals}) as an example, as the right-hand side of the sampling process for $ \mathbf{x}_{t - 1} $ contains no term that is relevant to itself, we can simply write the inverted sampling process that calculates $ \mathbf{x}_t $ as:
\begin{equation} 
    \label{eq:x_t_equals}
    \mathbf{x}_t = \frac{1 - \bar{\alpha}_t}{(1 - \bar{\alpha}_{t - 1}) \sqrt{\alpha_t}} (\mathbf{x}_{t - 1} - \sigma_t \mathbf{z}) - \frac{\beta_t \sqrt{\bar{\alpha}_{t - 1}}}{(1 - \bar{\alpha}_{t - 1}) \sqrt{\alpha_t}} \epsilon(\mathbf{x}_t, t).
\end{equation}
Here, we can ignore or cache the sampling term with randomness $ \sigma_t \mathbf{z} $ to make it constant.
Thus, the only obstacle to the inversion is the non-invertible denoising neural network $ \epsilon(\mathbf{x}_t, t) $ that requires $ \mathbf{x}_t $ to be known. 

\subsubsection{Inverted Denoising} 

To overcome this, the invertible diffusion generation process is proposed \cite{wallace2023edict}. Inspired from the Affine Coupling Layers (ACL) used in many normalizing flow models [?, ?], a new variable $ \mathbf{y}_t $ is introduced to form an affine couple that makes (\ref{eq:x_t_equals}) independent of $ \mathbf{x}_t $. With all above, if we rewrite (\ref{eq:x_t-1_equals}) as $ \mathbf{x}_{t - 1} = a_t \mathbf{x}_t + b_t \epsilon(\mathbf{x}_t, t) $, we now have:
\begin{equation}
    \mathbf{x}_{t - 1} = a_t \mathbf{x}_t + b_t \epsilon(\mathbf{y}_t, t), \ 
    \mathbf{y}_{t - 1} = a_t \mathbf{y}_t + b_t \epsilon(\mathbf{x}_{t - 1}, t).
\end{equation}
where initially $ \mathbf{y}_{T} = \mathbf{x}_{T} $. To recover $ \mathbf{x}_t $ from $ \mathbf{x}_{t - 1} $ and $ \mathbf{y}_{t - 1} $, we naturally have:
\begin{equation}
    \mathbf{y}_t = \frac{1}{a_t} \left( \mathbf{y}_{t - 1} - b_t \epsilon(\mathbf{x}_{t - 1}, t) \right), \ 
    \mathbf{x}_t = \frac{1}{a_t} \left( \mathbf{x}_{t - 1} - b_t \epsilon(\mathbf{y}_t, t) \right).
\end{equation}
After this, we see that all denoising and sampling processes of DDPM steps become invertible and we can now reconstruct the original noisy input $ \mathbf{x}_T $ from the generated result $ \mathbf{x}_0 $.
In practice, a mixing coefficient $ p \in [0, 1] $ is also designed to constrain $ \mathbf{x}_t $ and $ \mathbf{y}_t $ from diverging by reassigning the weighted average to the two variables \textit{simultaneously}: 
\begin{equation}
    \mathbf{x}_{t - 1} \leftarrow p \mathbf{x}_{t - 1} + (1 - p) \mathbf{y}_{t - 1}, \ 
    \mathbf{y}_{t - 1} \leftarrow (1 - p) \mathbf{x}_{t - 1} + p \mathbf{y}_{t - 1}.
\end{equation}

In summary, for a general diffusion model, we now process two variables $ \mathbf{x}_t $ and $ \mathbf{y}_t $ in parallel using the same original denoising neural network in each step instead of only one $ \mathbf{x}_t $, and produce two new samples $ \mathbf{x}_{t - 1} $ and $ \mathbf{y}_{t - 1} $ with slight differences if observed. These can again be inverted using the inverted sampling and denoising processes to reconstruct the original variables $ \mathbf{x}_t $ and $ \mathbf{y}_t $. The whole process does not require any modification to the original denoising neural network.

\subsection{Intermediate Noise Reconstruction} \label{sec:intermediate_noise_reconstruction}

\begin{figure}[t]
    \centering
    \includegraphics[scale=1.0]{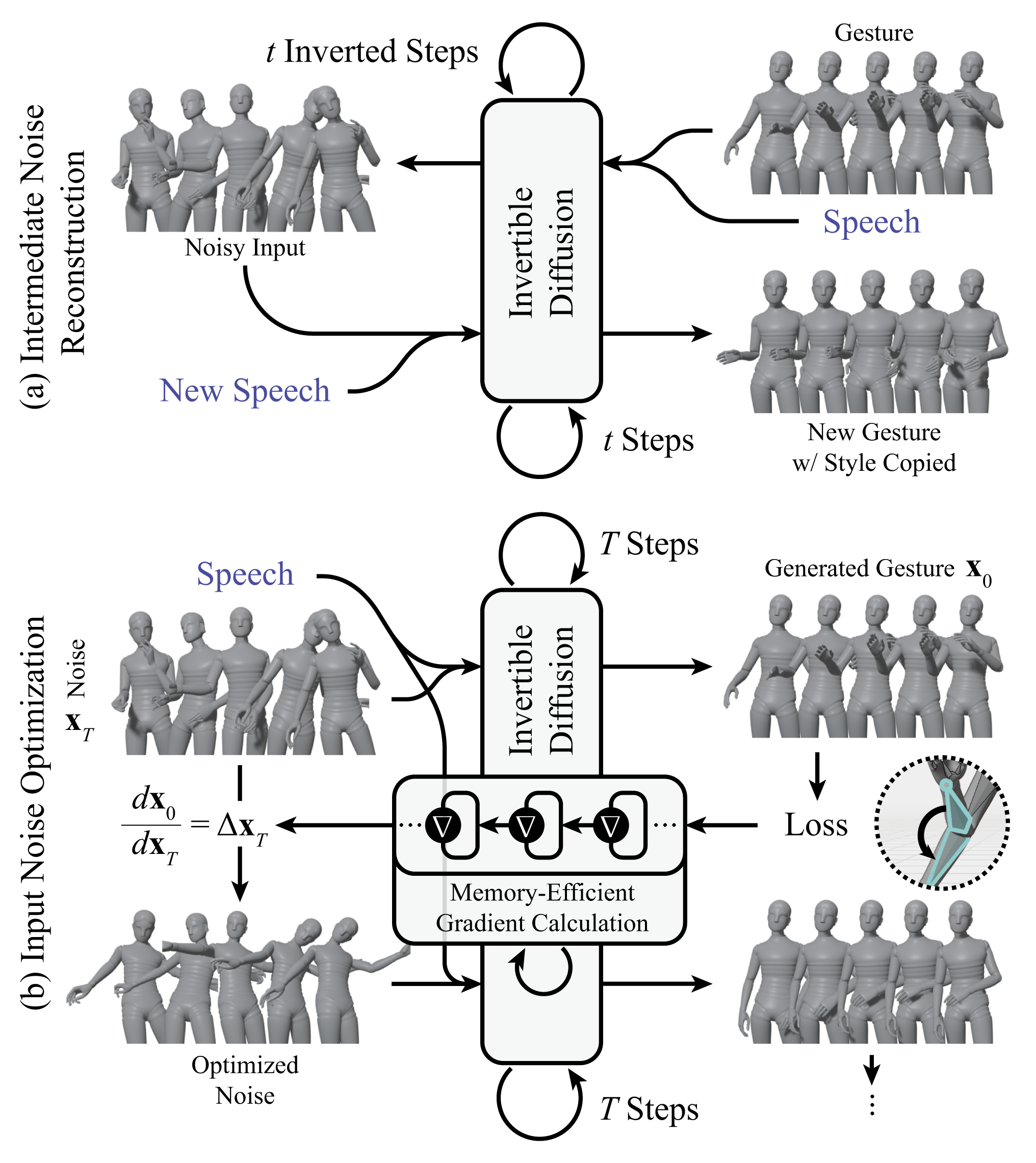}
    \caption{Two key capabilities of invertible diffusion models: (a) intermediate noise reconstruction for high-level editing and (b) input noise optimization for low-level editing.}
    \label{fig:keycapabilities}
\end{figure}

The majority of applications of diffusion models also come with the high-level editing capability utilizing condition variables $ C $. 
In these settings, the denoising neural network $ \epsilon(\mathbf{x}_t, t, C) $ takes $ C $ as an additional input, which often encodes guiding information like prompt text, speech audio, or image style.
With diffusion inversion, we can reconstruct arbitrary intermediate noisy samples $ \mathbf{x}_t $ of a generated result $ \mathbf{x}_0 $ that is ready for post-processing to output format with the introduction of the coupled $ \mathbf{y}_0 $.
Assuming $ \mathbf{y}_0 = \mathbf{x}_0 $, we can run this process to reconstruct the noisy input $ \mathbf{x}_t $ for any existing sample and regenerate with different conditions $ C^\prime $ for new results with strong high-level similarities to the existing samples, as shown in Fig. \ref{fig:keycapabilities} (a). 

Inspired by this, we utilize the intermediate noise reconstruction capability of the invertible diffusion models for co-speech gesture generation since diffusion models designed for the task take the speech audio, text, speaker identity, or other guiding information as conditions.
For an existing gesture sequence, this reconstruction-regeneration process can produce new gestures synchronized with new speech conditions with the style of the original gesture sequence well-preserved.
This can be seen as a style-copying editing technique for co-speech gesture generation that is not performed well by common diffusion models, as illustrated in Fig. \ref{fig:examples} (a).

Despite the theoretical feasibility, the limited representation precision of the floating-point numbers and converting losses of exported gesture results can cause severe numerical instability during the inversion process. 
The accumulated errors after thousands of inverted generation steps make the reconstructed noise incalculable in our practice.
To resolve this, we run the reconstruction process with a reduced total diffusion step count, e.g., from 1000 to 50.

\subsection{Input Noise Optimization} \label{sec:input_noise_optimization}

Directly optimizing input noise or latents can be a powerful editing method for generative models, as it directly moves the sampled point in the latent space in an automatic manner that enables guided control of the output without modification. 
As an example, DragGAN \cite{pan2023drag} leverages this to allow users to define pairs of source and target control points on an image and then regenerate with optimized input latents with tracking techniques to move the elements on the source position to the target position.
During the optimization, a loss function related to the paired points is defined and the gradients with respect to the latents are calculated for the stochastic gradient descent (SGD) process to minimize the loss.
However, for diffusion models, it is physically impossible for current hardware with limited memory to cache all activation states along the commonly thousands of denoising steps to get the value of $ \frac{d\mathbf{x}_0}{d\mathbf{x}_T} $ for optimization.
Direct Optimization of Diffusion Latents (DOODL) \cite{wallace2023end} is proposed to tackle this issue by utilizing diffusion inversion to make intermediate states between steps reconstructible, thus limiting the memory cost by reducing caching requirements (in our experiments from $ 70+ $ GB to $ 10- $ GB).
For this reason, diffusion inversion is considered to be the key factor that enables the direct optimization of input noises for diffusion models.
The schematic can be seen in Fig. \ref{fig:keycapabilities} (b).

Inspired by this, we argue that this input noise optimization capability of the invertible diffusion models can be ported to diffusion models designed for co-speech gesture generation task. 
Co-speech gestures are represented as sequences of frames with skeletons that consist of joint positions or rotations. 
It is much easier than images to define loss functions for joint rotations, velocity, or other low-level details with direct physical meaning.
After generating new gestures $ \mathbf{x}^{(0)}_0, \mathbf{y}^{(0)}_0 $, we calculate the gradients with respect to the input noise $ \mathbf{x}^{(0)}_T, \mathbf{y}^{(0)}_T $ and take an SGD step to get the new input noise $ \mathbf{x}^{(1)}_T, \mathbf{y}^{(1)}_T $ that minimizes the loss function $ L(\mathbf{x}_0) $.
Regenerate using the new input noise gives us the new gestures $ \mathbf{x}^{(1)}_0 $ that is expected to more conform to our editing goal.
These optimization steps are repeated until we get the satisfactory $ \mathbf{x}^{(s)}_0 $.
A loss function can be specifically designed for a variety of editing goals, e.g., specifying the rotation property of a specific joint in some frames, changing the velocity or range of the overall gesture motion, or making the left and right parts of the body more symmetrical, etc.
See examples in Fig. \ref{fig:examples} (b) and details in section \ref{sec:experiments}.

\section{EXPERIMENTS} \label{sec:experiments}

As stated in section \ref{sec:method}, diffusion inversion unifies high-level and low-level editing for co-speech gesture generation via its capabilities of intermediate noise reconstruction and input noise optimization.
To demonstrate its effectiveness, we design and conduct extensive experiments on multiple use cases with both subjective and objective evaluations.
We choose DiffuseStyleGesture+ \cite{yang2023diffusestylegesture+} as the base diffusion model, which is the best performing model in GENEA 2023 challenge \cite{kucherenko2023genea} in terms of appropriateness for agent speech among a number of diffusion models designed for the task.
For performance reasons, the model is trained on the data of speaker 2 and 10 from the BEAT \cite{liu2022beat} dataset, which contains a total of 289 minutes of speech audio, text, and synchronized motion captured gesture sequences.
After ported to an invertible version, the model takes speech audio, text, speaker identity, and seed gestures from the tail of the previous generated gesture sequence as conditions $ C $, and generates new gesture sequences $ \mathbf{x}_0, \mathbf{y}_0 $ from noisy input $ \mathbf{x}_T, \mathbf{y}_T $ in $ T = 1000 $ steps.

As many recent literatures \cite{nyatsanga2023comprehensive, kucherenko2023genea, ao2022rhythmic, alexanderson2023listen} suggested, subjective evaluations should be considered as the primary measure for the quality of co-speech gesture generation results.
To maximally reduce the difficulty of judgement for human evaluators under very limited conditions, the subjective evaluations based on human preferences are conducted in the manner of pairwise comparison questionnaires.
Nevertheless, we still provide objective evaluations using the commonly used FGD \cite{10.1145/3414685.3417838} metric which calculates the Fréchet inception distance of features extracted by an autoencoder \cite{liu2022beat} specifically trained on the selected BEAT dataset.
Lower scores indicate better similarity between the original and edited gestures.
Note that this metric is designed for comparisons of two distributions with large sample sizes and may not be accurate in our cases.
However, the relative magnitudes still provide useful information for the comparisons.


\subsection{Style-Preserving Regeneration} \label{sec:style_preserving_regeneration}

We first evaluate the high-level editing capability of diffusion inversion by a style-preserving regeneration task. 
As introduced in section \ref{sec:intermediate_noise_reconstruction}, this leverages the intermediate noise reconstruction capability to copy the style of an existing gesture to gestures for new speech conditions.
Specifically, we take an existing gesture sequence $ \mathbf{x}_0 $, define its coupled $ \mathbf{y}_0 = \mathbf{x}_0 $, and run the inversion process to reconstruct the noisy input $ \mathbf{x}_{1000} $, which, as mentioned, should be completed in 50 steps instead of 1000.
Then, we regenerate new gestures $ \mathbf{x}^\prime_0 $ with new speech conditions $ C^\prime $ from the reconstructed noisy input $ \mathbf{x}_{1000} $.
This can be completed with the full 1000 steps without problem.

\begin{figure}[tbp]
    \centering
    \includegraphics[scale=1.0]{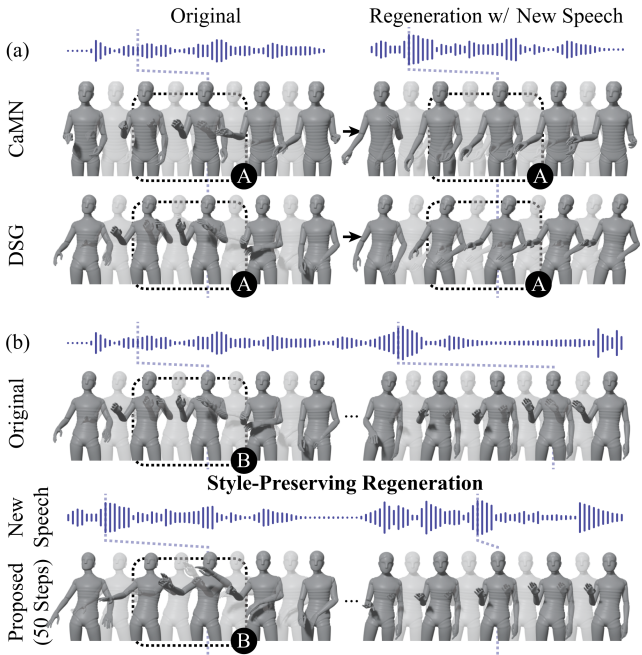}
    \caption{Demonstration of style-preserving regeneration. Compared to the baselines that produce much varied results for new speech with the same conditions (A), the proposed model gives more similar results to the original gestures in style (B). Input text is omitted. \textit{Zoom in} for closer look and more intermediate frames. \textbf{Same for Fig. \ref{fig:frame_joint_editing}, \ref{fig:motion_range_editing}, \ref{fig:velocity_editing}, and \ref{fig:symmetry_editing}.}}
    \label{fig:style_preserving_regeneration}
\end{figure}

For comparison, we also run similar tasks with two baseline models. First, we view the generated gestures of a non-invertible version of the base model (labeled DSG) \cite{yang2023diffusestylegesture+} as the existing gestures and generate for new speeches with the same speaker identity as we assume we cannot reconstruct the noisy input in this case.
Second, we perform the same process to the model CaMN \cite{liu2022beat} provided by the authors of the BEAT dataset with the speaker identity and emotion label unchanged during regeneration.
A total number of 60 questions are presented to 10 human evaluators, with each question asking about which of the two regenerated gesture sequences from one baseline and the proposed method is more similar to the existing gesture in style. 
Another 60 questions asking about which one has better human-likeness after regeneration are also presented, along with 60 more questions for synchronization with the input speech.
These questions are to ensure there is no severe degradation of regeneration quality.
The results are shown in Table \ref{tab:style_preserving_regeneration} along with the FGD scores.
See Fig. \ref{fig:style_preserving_regeneration} for demonstration.

\begin{table}[t]
\caption{Results of Style-Preserving Regeneration}
\label{tab:style_preserving_regeneration}
\begin{center}
\begin{tabular}{p{4em}rrrr}
\toprule
\multirow{4}{*}{Method} & \multirow{4}{*}{FGD $ \downarrow $} & \multicolumn{3}{c}{Human Preference (Win Rate \%)} \\
\cmidrule(l){3-5}
& & \begin{tabular}[c]{@{}c@{}} Style \\ Preservance \end{tabular} $ \uparrow $ & \begin{tabular}[c]{@{}c@{}} Post-Editing \\ Human-Likeness \end{tabular} $ \uparrow $ & \begin{tabular}[c]{@{}c@{}} P.-E. \\ Sync. \end{tabular} $ \uparrow $ \\
\midrule
Baselines \\ \\[-6pt]
CaMN & 879.8 & 18.3 & 30.0 & \textbf{35.0} \\
DSG & 1199.6 & 20.0 & \textbf{38.0} & 30.0 \\
\midrule
Proposed \\ \\[-6pt]
50 Steps & \textbf{155.6} & $ ^\ast $ \textbf{61.7} & 31.7 & \textbf{35.0} \\
\bottomrule
\end{tabular}
\end{center}
\footnotesize{$ ^\ast $ This indicates that for a total of 60 pairwise comparison questions asked (20 per case of 3 cases: CaMN vs. DSG, CaMN vs. 50 Steps, and DSG vs. 50 Steps), 61.7\%  answers prefer the proposed method with 50 Steps over the other. \textbf{Same for Table \ref{tab:frame_joint_editing}, \ref{tab:motion_range_editing}, \ref{tab:velocity_editing}, and \ref{tab:symmetry_editing}.}}
\end{table}

\subsection{Frame-Joint Editing}

We then evaluate the low-level editing capability of diffusion inversion by multiple tasks, first of which is the frame-joint editing task.
As mentioned in section \ref{sec:input_noise_optimization}, the input noise optimization capability enables us to directly optimize the input noise with the guidance of loss functions that directly operate on the low-level details of the generated gestures.
For this task, our aim is to regenerate gesture sequences with dedicated joint rotations $ \textbf{j} \in \mathbb{R}^{J^\prime \times R} $ in specific frames $ \textbf{f} \in \mathbb{N}^{F^\prime} $.
To do this, we first generate a gesture sequence $ \mathbf{x}_0 \in \mathbb{R}^{F \times J \times R} $ and calculate the gradients with respect to the reconstructed input noise $ \mathbf{x}_{1000} $ to minimize the frame-joint loss function using SGD:
\begin{equation}
    L_{\text{fj}}(\mathbf{x}_0) = \frac{1}{F^\prime J^\prime R} \sum_{f \in \textbf{f}} \sum_{j = 1}^{J^\prime} \sum_{r = 1}^{R} (\mathbf{x}_0[f, \indexfunc(j), r] - \textbf{j}[j, r])^2,
\end{equation}
where $ \indexfunc(j) $ is the index of the corresponding joint in the full gesture sequence.

\begin{figure}[tbp]
    \centering
    \includegraphics[scale=1.0]{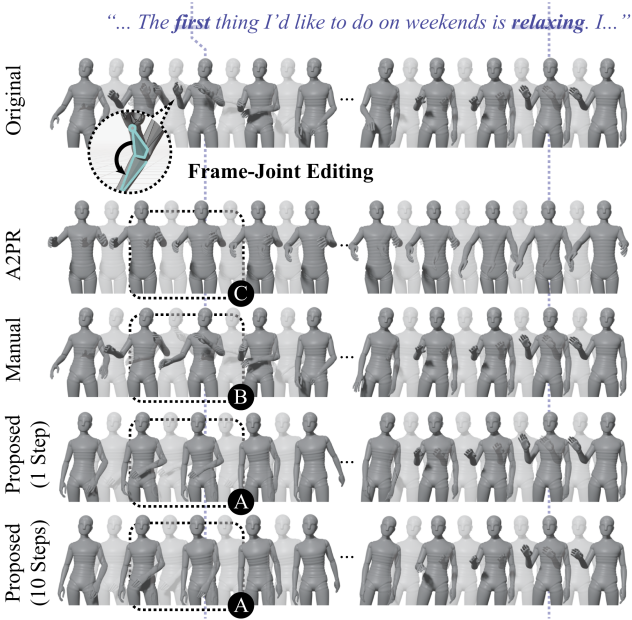}
    \caption{Demonstration of frame-joint editing. The proposed method can complete the editing goal with modifications better merged into the original gestures (A), instead of a less human-like interpolation when edited manually (B). The baseline method does not perform as well due to low editing resolution (one key frame per second) and skeleton definition conversion (C). Input audio is omitted.}
    \label{fig:frame_joint_editing}
\end{figure}

For comparison, we provide a baseline model Audio2PhotoReal \cite{ng2024audio} (labeled A2PR) that provides a pose diffusion model for producing gesture sequences from low frame rate guide poses. 
These guide poses can act like key frames for the full sequence and can be edited to influence the final results.
We perform similar editing to ours by manually tweaking the guide poses extracted from the original gesture sequences as required. 
Besides, we also compare the results to manually edited gestures (labeled Manual) as another baseline, with selected joint rotations in specific frames manually modified and nearby frames interpolated.
Two settings of the proposed method are evaluated, one with 1 step of optimization and the other with 10 steps.
In total, 120 questions are asked to the 10 human evaluators about which one of two models or settings better completes the editing goal, along with another 120 questions asking about human-likeness and another 120 for speech-gesture synchronization.
See results in Table \ref{tab:frame_joint_editing} and demonstration in Fig. \ref{fig:frame_joint_editing}.

\begin{table}[t]
\caption{Results of Frame-Joint Editing}
\label{tab:frame_joint_editing}
\begin{center}
\begin{tabular}{p{3.5em}rrrr}
\toprule
\multirow{4}{*}{Method} & \multirow{4}{*}{FGD $ \downarrow $} & \multicolumn{3}{c}{Human Preference (Win Rate \%)} \\
\cmidrule(l){3-5}
& & \begin{tabular}[c]{@{}c@{}} Editing Goal \\ Completion \end{tabular} $ \uparrow $ & \begin{tabular}[c]{@{}c@{}} Post-Editing \\ Human-Likeness \end{tabular} $ \uparrow $ & \begin{tabular}[c]{@{}c@{}} P.-E. \\ Sync. \end{tabular} $ \uparrow $ \\
\midrule
Baselines \\ \\[-6pt]
A2PR & 177.3 & 9.1 & 21.7 & 23.3 \\
Manual & 83.2 & 20.0 & 21.7 & 25.0 \\
\midrule
Proposed \\ \\[-6pt]
1 Step & \textbf{76.1} & \textbf{37.5} & \textbf{30.8} & 25.0 \\
50 Steps & 92.0 & 33.3 & 25.8 & \textbf{26.7} \\
\bottomrule
\end{tabular}
\end{center}
\end{table}

\subsection{Motion Range Editing} \label{sec:motion_range_editing}

Another low-level editing task we evaluate is motion range editing, where we aim to control the overall spatial range of arms, legs, etc. throughout the regenerated gesture sequence.
This can be done by defining the motion range loss function making the Euler angle rotations of every joint closer to zero:
\begin{equation}
    L_{\text{mr}}(\mathbf{x}_0) = \frac{1}{F J R} \sum_{f = 1}^F \sum_{j = 1}^J \sum_{r = 1}^R (\euler(\mathbf{x}_0)[f, j, r] - 0)^2.
\end{equation}
Minimizing such loss function can effectively reduce the rotation angles thus the motion range.
Increasing motion range can be done by minimizing the negative of the loss function.

\begin{figure}[tbp]
    \centering
    \includegraphics[scale=1.0]{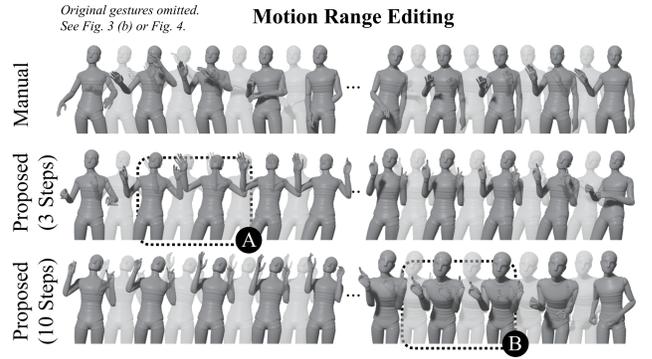}
    \caption{Demonstration of motion range editing. The proposed method regenerates more significant motion range changes compared to the manual editing (A). However more optimization steps does not necessarily lead to better results (B).}
    \label{fig:motion_range_editing}
\end{figure}

Similarly, we compare the results to manually edited gestures where all rotation angles are multiplied by a constant factor 2. 
Similarly, $ 3 \times 60 $ questions are asked about goal completion, human-likeness, and synchronization. 
The results are shown in Table \ref{tab:motion_range_editing}.
Also, see Fig. \ref{fig:motion_range_editing} for demonstration.

\begin{table}[t]
\caption{Results of Motion Range Editing}
\label{tab:motion_range_editing}
\begin{center}
\begin{tabular}{p{3.5em}rrrr}
\toprule
\multirow{4}{*}{Method} & \multirow{4}{*}{FGD $ \downarrow $} & \multicolumn{3}{c}{Human Preference (Win Rate \%)} \\
\cmidrule(l){3-5}
& & \begin{tabular}[c]{@{}c@{}} Editing Goal \\ Completion \end{tabular} $ \uparrow $ & \begin{tabular}[c]{@{}c@{}} Post-Editing \\ Human-Likeness \end{tabular} $ \uparrow $ & \begin{tabular}[c]{@{}c@{}} P.-E. \\ Sync. \end{tabular} $ \uparrow $ \\
\midrule
Baseline \\ \\[-6pt]
Manual & 983.8 & 8.3 & \textbf{50.0} & 33.3 \\
\midrule
Proposed \\ \\[-6pt]
3 Steps & \textbf{228.3} & 43.3 & 35.0 & \textbf{36.7} \\
10 Steps & 1199.8 & \textbf{48.8} & 15.0 & 30.0 \\
\bottomrule
\end{tabular}
\end{center}
\end{table}

\subsection{Velocity Editing}

\begin{figure}[tbp]
    \centering
    \includegraphics[scale=1.0]{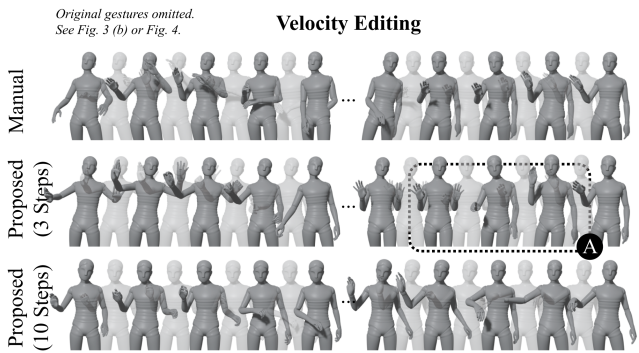}
    \caption{Demonstration of velocity editing. The edited gestures by the proposed method come with more small movements within the same length of time (A).}
    \label{fig:velocity_editing}
\end{figure}

Rotation velocity, defined here as the change rate of the rotation of the joints, can effect the overall motion speed of the generated gesture.
We define the velocity loss function similarly to the motion range loss:
\begin{equation}
    L_{\text{v}}(\mathbf{x}_0) = \frac{1}{F J R} \sum_{f = 1}^F \sum_{j = 1}^J \sum_{r = 1}^R (\vel(\mathbf{x}_0)[f, j, r] - 0)^2,
\end{equation}
where $ \vel(\mathbf{x}_0) $ calculates the inter-frame velocity by differentiating the joint rotation values between frames.
Similarly, minimizing the loss function or the negative of it can reduce or increase the overall motion speed of the regenerated gesture.
The results are also compared to manually edited gestures where the rotations are first converted to velocities, then multiplied by a constant factor 2, and finally integrated back to rotations. 
See evaluation (same as in section \ref{sec:motion_range_editing}) results in Table \ref{tab:motion_range_editing} and demonstration in Fig. \ref{fig:velocity_editing}.

\begin{table}[t]
\caption{Results of Velocity Editing}
\label{tab:velocity_editing}
\begin{center}
\begin{tabular}{p{3.5em}rrrr}
\toprule
\multirow{4}{*}{Method} & \multirow{4}{*}{FGD $ \downarrow $} & \multicolumn{3}{c}{Human Preference (Win Rate \%)} \\
\cmidrule(l){3-5}
& & \begin{tabular}[c]{@{}c@{}} Editing Goal \\ Completion \end{tabular} $ \uparrow $ & \begin{tabular}[c]{@{}c@{}} Post-Editing \\ Human-Likeness \end{tabular} $ \uparrow $ & \begin{tabular}[c]{@{}c@{}} P.-E. \\ Sync. \end{tabular} $ \uparrow $ \\
\midrule
Baseline \\ \\[-6pt]
Manual & 1230.8 & 25.0 & 6.7 & \textbf{35.0} \\
\midrule
Proposed \\ \\[-6pt]
3 Steps & \textbf{148.2} & \textbf{43.3} & \textbf{51.7} & 31.7 \\
10 Steps & 656.4 & 31.7 & 41.7 & 33.3 \\
\bottomrule
\end{tabular}
\end{center}
\end{table}

\subsection{Symmetry Editing}

\begin{figure}[tbp]
    \centering
    \includegraphics[scale=1.0]{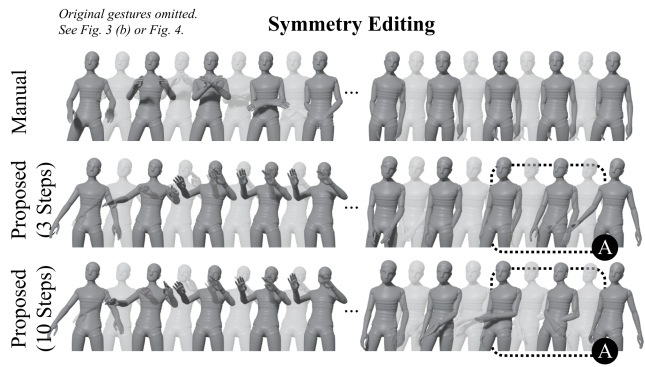}
    \caption{Demonstration of symmetry editing. Overall, the proposed method regenerates symmetrical gestures with much better human-likeness. This is another example of degraded results with more optimization steps (A).}
    \label{fig:symmetry_editing}
\end{figure}

Symmetry between the left and right parts of the upper body can also effect the overall feeling of the generated gesture as it in some sense reflects the speaker's confidence and emotion.
The symmetry loss function is thus defined comparing the left part $ \leftfunc(\mathbf{x}_0) $ and right part $ \rightfunc(\mathbf{x}_0) $ of the upper body:
\begin{equation}
    L_{\text{v}}(\mathbf{x}_0) = \frac{1}{F J R} \sum_{f = 1}^F \sum_{j = 1}^J \sum_{r = 1}^R (\leftfunc(\mathbf{x}_0)[f, j, r] - \rightfunc(\mathbf{x}_0)[f, j, r])^2.
\end{equation}
See demonstration in Fig. \ref{fig:symmetry_editing}, and evaluation (same as in section \ref{sec:motion_range_editing}) results in Table \ref{tab:symmetry_editing} where the regenerated results are compared to the manually edited gestures where the right part is mirrored to the left.

\begin{table}[t]
\caption{Results of Symmetry Editing}
\label{tab:symmetry_editing}
\begin{center}
\begin{tabular}{p{3.5em}rrrr}
\toprule
\multirow{4}{*}{Method} & \multirow{4}{*}{FGD $ \downarrow $} & \multicolumn{3}{c}{Human Preference (Win Rate \%)} \\
\cmidrule(l){3-5}
& & \begin{tabular}[c]{@{}c@{}} Editing Goal \\ Completion \end{tabular} $ \uparrow $ & \begin{tabular}[c]{@{}c@{}} Post-Editing \\ Human-Likeness \end{tabular} $ \uparrow $ & \begin{tabular}[c]{@{}c@{}} P.-E. \\ Sync. \end{tabular} $ \uparrow $ \\
\midrule
Baseline \\ \\[-6pt]
Manual & 424.9 & 26.7 & 10.0 & \textbf{35.0} \\
\midrule
Proposed \\ \\[-6pt]
3 Steps & \textbf{74.5} & \textbf{41.7} & \textbf{51.7} & 33.3 \\
10 Steps & 80.0 & 31.7 & 38.3 & 31.7 \\
\bottomrule
\end{tabular}
\end{center}
\end{table}

\subsection{Editing Latency} \label{sec:editing_latency}

\begin{figure}[t]
    \centering
    \subfloat[]{%
        \begin{tikzpicture}
            \begin{axis}[
                xlabel={Step},
                ylabel={Loss (Relative to Step 0)},
                xmin=0, xmax=10,
                ymin=0, ymax=1.0,
                xtick={0, 1, 2, 3, 4, 5, 6, 7, 8, 9, 10},
                ytick={0.2, 0.4, 0.6, 0.8, 1.0},
                legend pos=south east,
                ymajorgrids=true,
                grid style=dashed,
                scale=0.5,
                label style={font=\footnotesize},
                title style={font=\footnotesize},
                tick label style={font=\footnotesize},
                legend style={font=\footnotesize},
                x label style={at={(axis description cs:0.5, 0.075)}, anchor=north},
                y label style={at={(axis description cs:0.65, 1.025)}, rotate=270, anchor=south},
                y tick label style={/pgf/number format/.cd, fixed, fixed zerofill, precision=1, /tikz/.cd},
                legend style={fill opacity=0.75, text opacity=1}
            ]
            
            \addplot[color=blue, mark=square] coordinates {
                (0,  1.0)
                (1,  0.112878955)
                (2,  0.107313574)
                (3,  0.105175797)
                (4,  0.103918626)
                (5,  0.102114793)
                (6,  0.101601416)
                (7,  0.101324899)
                (8,  0.100182819)
                (9,  0.100363451)
                (10, 0.099460001)
            };
            \addlegendentry{$L_{\text{fj}}$}

            \addplot[color=red, mark=o] coordinates {
                (0,  1.0)
                (1,  0.911494779)
                (2,  0.842141547)
                (3,  0.754431005)
                (4,  0.726278841)
                (5,  0.717215195)
                (6,  0.70763012 )
                (7,  0.707480795)
                (8,  0.69599806 )
                (9,  0.694025336)
                (10, 0.68723789 )
            };
            \addlegendentry{$L_{\text{mr}}$}

            \addplot[color=cyan, mark=triangle] coordinates {
                (0,  1.0)
                (1,  0.985522534)
                (2,  0.604720213)
                (3,  0.575841477)
                (4,  0.549845143)
                (5,  0.550266022)
                (6,  0.55469614 )
                (7,  0.532947516)
                (8,  0.512148814)
                (9,  0.525012067)
                (10, 0.499977367)
            };
            \addlegendentry{$L_{\text{v}}$}

            \addplot[color=magenta, mark=pentagon] coordinates {
                (0,  1.0)
                (1,  0.853661087)
                (2,  0.787458117)
                (3,  0.775797535)
                (4,  0.764332826)
                (5,  0.758045464)
                (6,  0.754783364)
                (7,  0.750341484)
                (8,  0.747813872)
                (9,  0.744102496)
                (10, 0.740873042)
            };
            \addlegendentry{$L_{\text{s}}$}

            \end{axis}
        \end{tikzpicture}
    }
    \subfloat[]{%
        \begin{tikzpicture}
            \begin{axis}[
                xlabel={System},
                ylabel={Execution Time (s)},
                symbolic x coords={DOODL, T2M$ ^\ast $, Proposed},
                ytick={1e2, 2e2, 3e2, 5e2, 1e3},
                legend pos=north east,
                ymajorgrids=true,
                grid style=dashed,
                scale=0.5,
                label style={font=\footnotesize},
                title style={font=\footnotesize},
                tick label style={font=\footnotesize},
                legend style={font=\footnotesize},
                x label style={at={(axis description cs:0.5, 0.075)}, anchor=north},
                y label style={at={(axis description cs:0.65, 1.025)}, rotate=270, anchor=south},
                legend style={fill opacity=0.75, text opacity=1},
                ybar=true,
                enlarge x limits=0.3,
                enlarge y limits={value=0.2, upper},
                xtick align=inside,
                scaled ticks=false,
                nodes near coords,
                nodes near coords style={font=\tiny, color=black, yshift=-1},
            ]
            
            \addplot coordinates {
                (DOODL, 1003)
                (T2M$ ^\ast $, 300)
                (Proposed, 14)
            };
            
            \addplot coordinates {
                (DOODL, 252)
                (T2M$ ^\ast $, 180)
                (Proposed, 5)
            };

            \legend{Typical, Lowest}

            \end{axis}
        \end{tikzpicture}
    }
    \caption{Plots of (a) typical dropping curves of different guiding losses, and (b) execution time comparison of different systems utilizing diffusion inversion. $ ^\ast $Data from \cite{karunratanakul2023optimizing}.}
    \label{fig:loss_drop}
 \end{figure}
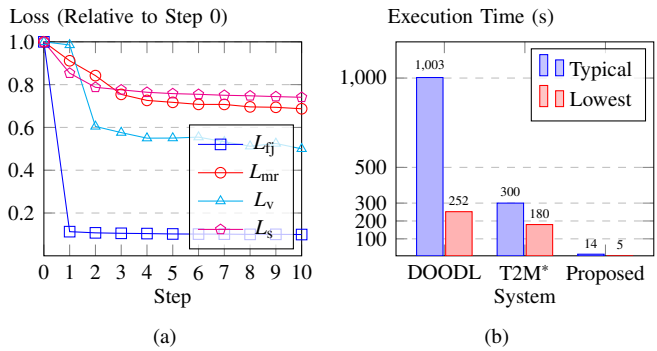

For content creators as the end users of such editing systems for co-speech gesture generation, the latency from when the task is specified to when the final results can be obtained is an important factor to build a reliable and practical application.
Both intermediate noise reconstruction and input noise optimization are designed to be completed in a number of steps, which challenges the time efficiency of the method.
As mentioned, the intermediate noise reconstruction can be completed in a reduced number of inverted diffusion generation steps of 50, which in our practice can be finished in $ \sim\!500 $ ms on our system with an RTX 4090 GPU.
Even with the $ \sim\!4000 $ ms for the full 1000 regeneration steps added, 9 out of 10 human evaluators agree that the latency of intermediate noise reconstruction applications is totally acceptable.
For input noise optimization applications, we find that in most cases the loss function can be reduced to a satisfactory level in less than 3 steps, as shown in Fig. \ref{fig:loss_drop} (a). 
Evaluation results in above subsections also suggest that optimizing for 10 steps does not significantly improve or even degrades the final results.
8 out of 10 human evaluators agree that less-than-15-second latency of input noise optimization applications is acceptable.
Compared to other editing systems utilizing diffusion inversion for image generation (DOODL \cite{wallace2023end}) and text-to-motion generation (T2M \cite{karunratanakul2023optimizing}), the execution time of our application is far more reasonable and practical for real content creators that prefer near real-time editing experiences, as shown in Fig. \ref{fig:loss_drop} (b).
With less complex diffusion models for co-speech gesture generation, compared to the other two, our application is much more suitable for practical use cases.

\section{CONCLUSIONS}

To provide an easy editing measure for diffusion-based co-speech gesture generation, we utilize diffusion inversion as a method that unifies both high-level and low-level editing tasks without re-training the diffusion model itself. 
We demonstrate its effectiveness by extensive quantitative and qualitative evaluations with five specialized editing tasks.
The results show that the proposed method can effectively complete the editing tasks with high level of gesture quality, editing goal completion, and human-likeness with acceptable latency.

However, this method still suffers from limitations due to its design and usage. 
As no modification or re-training are required for the diffusion model, the quality of the generated gestures depends heavily on the original model's performance.
If the original model lacks the capabilities to provide diverse results from the same noise but different conditions, tasks like the style-preserving regeneration will not be able to generate gestures with distinctive details.
Also, if too many optimization steps are taken for an editing task where the input noise is guided outside the original safe zone in the sampling space, the quality of the generated gestures may greatly degrade.
If the guiding loss function is not well-designed or overcomplicated, the optimization process will not be guaranteed to converge to a satisfactory result.
Possible improvements regarding these limitations can be studied in future works.










\bibliographystyle{IEEEtran}
\bibliography{IEEEabrv, PAPER}

\begin{thebibliography}{10}
\providecommand{\url}[1]{#1}
\csname url@rmstyle\endcsname
\providecommand{\newblock}{\relax}
\providecommand{\bibinfo}[2]{#2}
\providecommand\BIBentrySTDinterwordspacing{\spaceskip=0pt\relax}
\providecommand\BIBentryALTinterwordstretchfactor{4}
\providecommand\BIBentryALTinterwordspacing{\spaceskip=\fontdimen2\font plus
\BIBentryALTinterwordstretchfactor\fontdimen3\font minus \fontdimen4\font\relax}
\providecommand\BIBforeignlanguage[2]{{%
\expandafter\ifx\csname l@#1\endcsname\relax
\typeout{** WARNING: IEEEtran.bst: No hyphenation pattern has been}%
\typeout{** loaded for the language `#1'. Using the pattern for}%
\typeout{** the default language instead.}%
\else
\language=\csname l@#1\endcsname
\fi
#2}}

\bibitem{zhu2023taming}
L.~Zhu, X.~Liu, X.~Liu, R.~Qian, Z.~Liu, and L.~Yu, ``Taming diffusion models for audio-driven co-speech gesture generation,'' \emph{arXiv preprint arXiv:2303.09119}, 2023.

\bibitem{alexanderson2023listen}
S.~Alexanderson, R.~Nagy, J.~Beskow, and G.~E. Henter, ``Listen, denoise, action! audio-driven motion synthesis with diffusion models,'' \emph{ACM Transactions on Graphics (TOG)}, vol.~42, no.~4, pp. 1--20, 2023.

\bibitem{ao2023gesturediffuclip}
T.~Ao, Z.~Zhang, and L.~Liu, ``Gesturediffuclip: Gesture diffusion model with clip latents,'' \emph{ACM Transactions on Graphics (TOG)}, vol.~42, no.~4, pp. 1--18, 2023.

\bibitem{yang2023diffusestylegesture+}
S.~Yang, H.~Xue, Z.~Zhang, M.~Li, Z.~Wu, X.~Wu, S.~Xu, and Z.~Dai, ``The diffusestylegesture+ entry to the genea challenge 2023,'' in \emph{Proceedings of the 25th International Conference on Multimodal Interaction}, 2023, pp. 779--785.

\bibitem{ng2024audio}
E.~Ng, J.~Romero, T.~Bagautdinov, S.~Bai, T.~Darrell, A.~Kanazawa, and A.~Richard, ``From audio to photoreal embodiment: Synthesizing humans in conversations,'' \emph{arXiv preprint arXiv:2401.01885}, 2024.

\bibitem{10.1145/3414685.3417838}
Y.~Yoon, B.~Cha, J.-H. Lee, M.~Jang, J.~Lee, J.~Kim, and G.~Lee, ``Speech gesture generation from the trimodal context of text, audio, and speaker identity,'' \emph{ACM Trans. Graph.}, vol.~39, no.~6, pp. 1--16, nov 2020.

\bibitem{liu2022beat}
H.~Liu, Z.~Zhu, N.~Iwamoto, Y.~Peng, Z.~Li, Y.~Zhou, E.~Bozkurt, and B.~Zheng, ``Beat: A large-scale semantic and emotional multi-modal dataset for conversational gestures synthesis,'' in \emph{European conference on computer vision}.\hskip 1em plus 0.5em minus 0.4em\relax Springer, 2022, pp. 612--630.

\bibitem{wallace2023edict}
B.~Wallace, A.~Gokul, and N.~Naik, ``Edict: Exact diffusion inversion via coupled transformations,'' in \emph{Proceedings of the IEEE/CVF Conference on Computer Vision and Pattern Recognition}, 2023, pp. 22\,532--22\,541.

\bibitem{10.1145/2485895.2485900}
S.~Marsella, Y.~Xu, M.~Lhommet, A.~Feng, S.~Scherer, and A.~Shapiro, ``Virtual character performance from speech,'' in \emph{Proceedings of the 12th ACM SIGGRAPH/Eurographics Symposium on Computer Animation}, ser. SCA '13.\hskip 1em plus 0.5em minus 0.4em\relax New York, NY, USA: Association for Computing Machinery, 2013, pp. 25--35.

\bibitem{10.1145/1778765.1778861}
\BIBentryALTinterwordspacing
S.~Levine, P.~Kr{\"a}henb{\"u}hl, S.~Thrun, and V.~Koltun, ``Gesture controllers,'' \emph{ACM Trans. Graph.}, vol.~29, no.~4, pp. 1--11, jul 2010. [Online]. Available: \url{https://doi.org/10.1145/1778765.1778861}
\BIBentrySTDinterwordspacing

\bibitem{nyatsanga2023comprehensive}
S.~Nyatsanga, T.~Kucherenko, C.~Ahuja, G.~E. Henter, and M.~Neff, ``A comprehensive review of data-driven co-speech gesture generation,'' \emph{arXiv preprint arXiv:2301.05339}, 2023.

\bibitem{ginosar2019learning}
S.~Ginosar, A.~Bar, G.~Kohavi, C.~Chan, A.~Owens, and J.~Malik, ``Learning individual styles of conversational gesture,'' in \emph{Proceedings of the IEEE/CVF Conference on Computer Vision and Pattern Recognition}, Long Beach, CA, USA, 2019, pp. 3497--3506.

\bibitem{li2021audio2gestures}
J.~Li, D.~Kang, W.~Pei, X.~Zhe, Y.~Zhang, Z.~He, and L.~Bao, ``Audio2gestures: Generating diverse gestures from speech audio with conditional variational autoencoders,'' in \emph{Proceedings of the IEEE/CVF International Conference on Computer Vision}, 2021, pp. 11\,293--11\,302.

\bibitem{ye2022audio}
S.~Ye, Y.-H. Wen, Y.~Sun, Y.~He, Z.~Zhang, Y.~Wang, W.~He, and Y.-J. Liu, ``Audio-driven stylized gesture generation with flow-based model,'' in \emph{Computer Vision--ECCV 2022: 17th European Conference, Tel Aviv, Israel, October 23--27, 2022, Proceedings, Part V}.\hskip 1em plus 0.5em minus 0.4em\relax Springer, 2022, pp. 712--728.

\bibitem{liu2022learning}
X.~Liu, Q.~Wu, H.~Zhou, Y.~Xu, R.~Qian, X.~Lin, X.~Zhou, W.~Wu, B.~Dai, and B.~Zhou, ``Learning hierarchical cross-modal association for co-speech gesture generation,'' in \emph{Proceedings of the IEEE/CVF Conference on Computer Vision and Pattern Recognition}, 2022, pp. 10\,462--10\,472.

\bibitem{yang2023qpgesture}
S.~Yang, Z.~Wu, M.~Li, Z.~Zhang, L.~Hao, W.~Bao, and H.~Zhuang, ``Qpgesture: Quantization-based and phase-guided motion matching for natural speech-driven gesture generation,'' in \emph{Proceedings of the IEEE/CVF Conference on Computer Vision and Pattern Recognition}, 2023, pp. 2321--2330.

\bibitem{ao2022rhythmic}
T.~Ao, Q.~Gao, Y.~Lou, B.~Chen, and L.~Liu, ``Rhythmic gesticulator: Rhythm-aware co-speech gesture synthesis with hierarchical neural embeddings,'' \emph{ACM Transactions on Graphics (TOG)}, vol.~41, no.~6, pp. 1--19, 2022.

\bibitem{9898573}
Z.~Zhao, N.~Gao, Z.~Zeng, and S.~Zhang, ``Generating diverse gestures from speech using memory networks as dynamic dictionaries,'' in \emph{2022 International Conference on Culture-Oriented Science and Technology (CoST)}, 2022, pp. 163--168.

\bibitem{zhou2022gesturemaster}
C.~Zhou, T.~Bian, and K.~Chen, ``Gesturemaster: Graph-based speech-driven gesture generation,'' in \emph{Proceedings of the 2022 International Conference on Multimodal Interaction}, 2022, pp. 764--770.

\bibitem{zhou2022audio}
Y.~Zhou, J.~Yang, D.~Li, J.~Saito, D.~Aneja, and E.~Kalogerakis, ``Audio-driven neural gesture reenactment with video motion graphs,'' in \emph{Proceedings of the IEEE/CVF Conference on Computer Vision and Pattern Recognition}, 2022, pp. 3418--3428.

\bibitem{10.1145/3577190.3616118}
Z.~Zhao, N.~Gao, Z.~Zeng, G.~Zhang, J.~Liu, and S.~Zhang, ``Gesture motion graphs for few-shot speech-driven gesture reenactment,'' in \emph{Proceedings of the 25th International Conference on Multimodal Interaction}.\hskip 1em plus 0.5em minus 0.4em\relax New York, NY, USA: ACM, 2023, p. 772–778.

\bibitem{kucherenko2023genea}
T.~Kucherenko, R.~Nagy, Y.~Yoon, J.~Woo, T.~Nikolov, M.~Tsakov, and G.~E. Henter, ``The {GENEA} {C}hallenge 2023: {A} large-scale evaluation of gesture generation models in monadic and dyadic settings,'' in \emph{Proceedings of the ACM International Conference on Multimodal Interaction}, ser. ICMI '23.\hskip 1em plus 0.5em minus 0.4em\relax ACM, 2023.

\bibitem{mirza2014conditional}
M.~Mirza and S.~Osindero, ``Conditional generative adversarial nets,'' 2014.

\bibitem{hu2021lora}
E.~J. Hu, Y.~Shen, P.~Wallis, Z.~Allen-Zhu, Y.~Li, S.~Wang, L.~Wang, and W.~Chen, ``Lora: Low-rank adaptation of large language models,'' \emph{arXiv preprint arXiv:2106.09685}, 2021.

\bibitem{zhang2023adding}
L.~Zhang, A.~Rao, and M.~Agrawala, ``Adding conditional control to text-to-image diffusion models,'' in \emph{Proceedings of the IEEE/CVF International Conference on Computer Vision}, 2023, pp. 3836--3847.

\bibitem{pan2023drag}
X.~Pan, A.~Tewari, T.~Leimk{\"u}hler, L.~Liu, A.~Meka, and C.~Theobalt, ``Drag your gan: Interactive point-based manipulation on the generative image manifold,'' in \emph{ACM SIGGRAPH 2023 Conference Proceedings}, 2023, pp. 1--11.

\bibitem{wallace2023end}
B.~Wallace, A.~Gokul, S.~Ermon, and N.~Naik, ``End-to-end diffusion latent optimization improves classifier guidance,'' in \emph{Proceedings of the IEEE/CVF International Conference on Computer Vision}, 2023, pp. 7280--7290.

\bibitem{karunratanakul2023optimizing}
K.~Karunratanakul, K.~Preechakul, E.~Aksan, T.~Beeler, S.~Suwajanakorn, and S.~Tang, ``Optimizing diffusion noise can serve as universal motion priors,'' 2023.

\bibitem{sohl2015deep}
J.~Sohl-Dickstein, E.~Weiss, N.~Maheswaranathan, and S.~Ganguli, ``Deep unsupervised learning using nonequilibrium thermodynamics,'' in \emph{International conference on machine learning}.\hskip 1em plus 0.5em minus 0.4em\relax PMLR, 2015, pp. 2256--2265.

\bibitem{ho2020denoising}
J.~Ho, A.~Jain, and P.~Abbeel, ``Denoising diffusion probabilistic models,'' \emph{Advances in neural information processing systems}, vol.~33, pp. 6840--6851, 2020.

\bibitem{song2020denoising}
J.~Song, C.~Meng, and S.~Ermon, ``Denoising diffusion implicit models,'' \emph{arXiv preprint arXiv:2010.02502}, 2020.

\bibitem{hertz2022prompt}
A.~Hertz, R.~Mokady, J.~Tenenbaum, K.~Aberman, Y.~Pritch, and D.~Cohen-Or, ``Prompt-to-prompt image editing with cross attention control,'' \emph{arXiv preprint arXiv:2208.01626}, 2022.

\end{thebibliography}

\end{document}